\documentclass[conference]{IEEEtran}
\IEEEoverridecommandlockouts
\usepackage{cite}
\usepackage{xurl}
\usepackage{amsmath,amssymb,amsfonts}
 \usepackage[titlenumbered,ruled,noend]{algorithm2e}
\usepackage{tabularx}
\usepackage{graphicx}
\usepackage{textcomp}
\usepackage[driverfallback=dvipdfm]{hyperref}
\usepackage{xcolor}
\usepackage{booktabs}
\usepackage{multirow}
\usepackage{enumitem}   
\usepackage{cleveref}
\usepackage{lipsum}
\usepackage{listings}
\usepackage{stfloats}
\usepackage{twemojis}
\usepackage{float}
\usepackage[bottom]{footmisc}
\def\BibTeX{{\rm B\kern-.05em{\sc i\kern-.025em b}\kern-.08em
    T\kern-.1667em\lower.7ex\hbox{E}\kern-.125emX}}
\begin{document}

\title{Why is the User Interface a Dark Pattern? : Explainable Auto-Detection and its Analysis}
\author{
    \IEEEauthorblockN{
        Yuki Yada\IEEEauthorrefmark{1},
        Tsuneo Matsumoto\IEEEauthorrefmark{2}, 
        Fuyuko Kido\IEEEauthorrefmark{4},
        Hayato Yamana\IEEEauthorrefmark{1}
    }
    \IEEEauthorblockA{\IEEEauthorrefmark{1} \textit{Waseda University}, Tokyo, Japan, E-mail: \{yada\_yuki, yamana\}@yama.info.waseda.ac.jp}
    \IEEEauthorblockA{\IEEEauthorrefmark{2} \textit{National Consumer Affairs Center of Japan}, Sagamihara, Kanagawa, Japan}
    \IEEEauthorblockA{\IEEEauthorrefmark{4} \textit{Waseda Research Institute for Science and Engineering}, Tokyo, Japan, Email: fkido@aoni.waseda.jp}
}

\maketitle

\begin{abstract}
Dark patterns are deceptive user interface designs for online services that make users behave in unintended ways. Dark patterns, such as privacy invasion, financial loss, and emotional distress, can harm users. These issues have been the subject of considerable debate in recent years. In this paper, we study interpretable dark pattern auto-detection, that is, why a particular user interface is detected as having dark patterns. First, we trained a model using transformer-based pre-trained language models, BERT, on a text-based dataset for the automatic detection of dark patterns in e-commerce. Then, we applied post-hoc explanation techniques, including local interpretable model agnostic explanation (LIME) and Shapley additive explanations (SHAP), to the trained model, which revealed which terms influence each prediction as a dark pattern. In addition, we extracted and analyzed terms that affected the dark patterns. Our findings may prevent users from being manipulated by dark patterns, and aid in the construction of more equitable internet services. Our code is available at https://github.com/yamanalab/why-darkpattern.
\end{abstract}

\begin{IEEEkeywords}
Dark patterns, user protection, privacy, text classification, and explainable AI
\end{IEEEkeywords}

\section{Introduction}\label{sec:intro}

\subsection{Dark Pattern}
Dark patterns are malicious user interface designs that are cleverly crafted to lead end users to unintended or unexpected actions that benefit the service provider. Dark patterns have become a concern in recent years and various studies have been conducted on them, including classifying dark patterns, large-scale surveys on dark pattern-related internet services, and dark pattern auto-detection.

The term "Dark pattern" was first defined in 2010 by Brignull \cite{darkpattern.org}
through the website \url{darkpattern.org} as "tricks used in websites and apps that make a user do things that the user did not mean to, like buying or signing up for something." Prior studies have shown that dark patterns exist everywhere, including on e-commerce sites \cite{11kScale}, consenting to cookies \cite{cookieConsent}, online games \cite{gameDp}, and SNS \cite{mobileApp}.

In 2019, Mathur et al. conducted a large-scale study on dark patterns on e-commerce sites \cite{11kScale}. They then categorized the collected dark pattern instances into seven categories, as shown in Table \ref{tab:dptype:mathur}. Mathur et al.'s categories expanded on the dark pattern categories defined by Gray et al. \cite{darkSideUx} and Brignull \cite{darkpattern.org}. 

\begin{figure}[t]
\includegraphics[width=\linewidth]{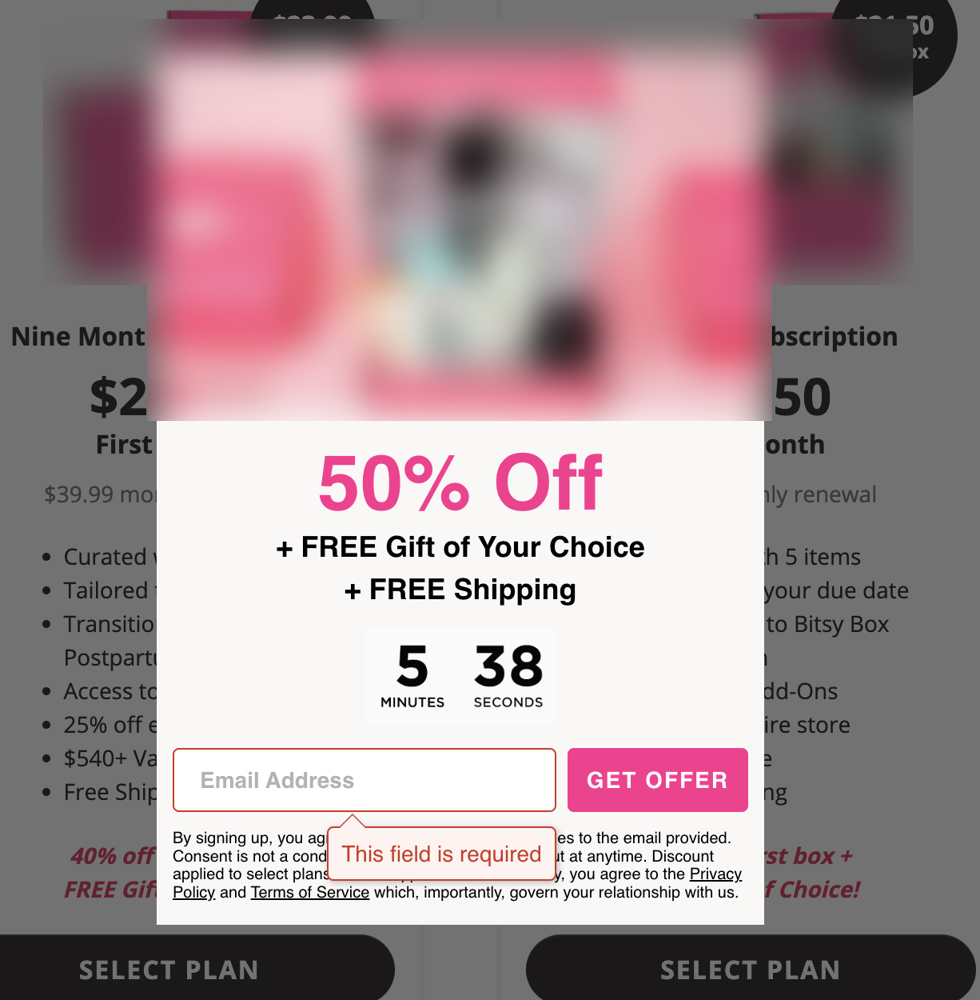}
\caption{Dark Patterns (Urgency) on \url{bumpboxes.com}. If revisited, the countdown timer resets; thus, there is no time limit}
\label{dpex}
\end{figure}

We provide a concrete example of a dark pattern on an e-commerce site in Fig. \ref{dpex}, which shows a dark pattern classified as ”urgency.” Urgency displays a short deadline for sale or purchase, giving users an undue sense of urgency and leading them to purchase the products. On this site, when users access the top page, they are shown a message offering a 50\% discount if they register as a member within a certain time, accompanied by a 6-minute countdown. However, if users revisit the page, the countdown timer that is in progress resets and a new 6-minute countdown began. Therefore, there is no deadline for the 50\% discount and the displayed countdown timer is deceptive \footnote{It should be noted that this case does not always mean that the site administrator intended to place a dark pattern. The timer might have run for 6 minutes to reserve the limited number of rights.
}. Dark patterns that give users a sense of urgency with ineffective countdowns, and lead them to accept discounts or offers, are known as urgency patterns.

\begin{table*}[t]
	\centering
	\caption{Types of Dark Patterns Defined by Mathur et al. \cite{11kScale}}
	\def\arraystretch{1.5}%  1 is the default, change whatever you need
	\begin{tabularx}{\textwidth}{l|X}  \hline
  Types of Dark Patterns & Description \\ \hline
        Forced Action   & Require users to take certain additional and tangential actions to complete their tasks \\ 
        Misdirection    &  Steer users toward or away from making a particular choice \\ 
        Obstruction & Make a certain action harder than it should be in order to dissuade users from taking that action. \\
        Scarcity    & Signal the limited availability or high demand of a product, thus increasing its perceived value and desirability \\ 
        Sneaking    & Attempt to misrepresent user actions, or hide/delay information that, if made available to users, they would likely object to\\ 
        Social Proof    & Manipulation of the user interface that privileges certain actions over others. \\ 
        Urgency &  Impose a deadline on a sale or deal, thereby accelerating user decision-making and purchases \\ \hline
	\end{tabularx}
	\label{tab:dptype:mathur}
\end{table*}

\subsection{Dark Pattern and User Protection}

In recent years, concerns over dark patterns, which are viewed as problematic regarding user protection, have increased in academic research and various areas, such as laws and regulations\cite{cpra,ftc,oecd,edpb}.

In 2020, the California Privacy Rights Act (CPRA) passed in California \cite{cpra}. The CPRA defines dark patterns as ”a user interface designed or manipulated with the substantial effect of subverting or impairing user autonomy, decision-making, or choice, as further defined by regulation” and discusses the potential impact of dark patterns on consumers.

In 2021, the Federal Trade Commission (FTC) \cite{ftc} issued a policy statement prohibiting subscription providers from trapping users to prevent them from canceling subscription agreements and using deceptive sign-up tactics. 

In 2022, the Organization for Economic Co-operation and Development (OECD) \cite{oecd} published a report on dark patterns (referred to as dark commercial patterns), based on discussions from a roundtable held in 2020. This report provides a practical definition of dark patterns, examples of their use, and their impact on users. It also presents policy proposals to assist policymakers. 

In 2022, the European Data Protection Board (EDPB) published guidelines for designers regarding dark patterns \cite{edpb}. The guidelines provide specific and categorized examples of dark patterns that might infringe upon users’ data protection rights and go against GDPR requirements intended to prevent such practices beforehand.

\subsection{Contributions}

Prior work \cite{11kScale,cookieConsent,gameDp,mobileApp} investigated dark patterns on e-commerce sites, consent to cookies, online games, and SNS, and attempted the automatic detection of dark patterns. However, to the best of our knowledge, no prior work has attempted to automatically detect dark patterns, except for our previous research \cite{dpInEc} and  Mansur et al. \cite{aidui}. However, no study has investigated why an automatically extracted pattern has a dark pattern.

Clarifying the reason why a given user interface is determined as a dark pattern is desirable because we can provide pieces of knowledge 1) to end-users for recognizing dark patterns and 2) to web designers for instructing new insights to design web pages, preventing unintentional implementation of dark patterns that designers have never known before.

In this study, we show how explainable machine learning models \cite{lime,shap} work well to explain why a given user interface is determined to have a dark pattern. We trained machine learning models using the E-Commerce Dark Pattern Dataset \cite{dpInEc}  to evaluate automatic dark pattern detection schemes. We employed transformer-based pre-trained models, including BERT \cite{bert} and RoBERTa \cite{roberta}. For the trained models, we adopted LIME \cite{lime} and SHAP \cite{shap}, which are post-hoc explanatory extraction methods. Subsequently, we analyzed the derived terms that influenced dark patterns.

\subsection{Organizations of this paper}

In the remainder of this paper, we present dark pattern taxonomies and definitions as preliminary knowledge in Section \ref{sec:pre}. Section \ref{sec:related} introduces related research on dark patterns. Section \ref{sec:method} describes the automatic detection of dark patterns on e-commerce sites using transformer-based pretrained language models, in addition to the explainability of extracted dart patterns using LIME and SHAP. Section \ref{sec:exp} presents our experimental results. In Section \ref{sec:ana}, we analyze the terms that influence the dark pattern decisions. Finally, we conclude this paper in Section \ref{sec:con}.

\section{Preliminary}\label{sec:pre}

\subsection{Dark Pattern Taxonomies and Definitions}

Various definitions exist for dark patterns. In 2010, Brignull \cite{darkpattern.org} introduced the term ”dark pattern” on his website darkpattern.org. He categorized these into 16 types and provided definitions and specific examples for each site.

In 2018, Gray et al. \cite{darkSideUx} created a new taxonomy of dark patterns consisting of five categories: Nagging, Obstruction, Sneaking, Interface Interference, and Forced Action. Two researchers, one expert in computer science and the other in UX design, studied dark patterns targeting online content such as search engines (Google and Bing) and social network services (Facebook, Twitter, and Reddit). Gray et al. revealed Brignull’s original taxonomy of dark patterns.

In 2021, Mathur et al. \cite{whatDp} conducted a comprehensive survey of dark patterns, targeting academic research, legislation, and regulations. As a result, through 11 literature sources, including academic presentations related to HCI and Privacy, and government documents such as legislation enacting regulations on dark patterns, they discovered that 84 types of dark patterns have been defined.

\section{Related Work}\label{sec:related}

To the best of our knowledge, no prior work has attempted to estimate why a given user interface has a dark pattern; thus, handcrafted and automatic detection of dark patterns is introduced below.

\subsection{Dark Patterns at Scale}

Many researchers have conducted extensive studies on dark patterns targeting various internet services and have confirmed the presence of dark patterns everywhere.

In 2019, Mathur et al. \cite{11kScale} conducted a large-scale survey of dark patterns on 11 K e-commerce sites. They extracted popular e-commerce sites using Alexa Traffic Rank. They manually investigated the product pages on these e-commerce sites to determine the presence of dark patterns. As a result of their investigation, dark patterns were identified in 11.1\% of e-commerce sites, which equates to 1,254 sites, with a total of 1,818 dark patterns.

In 2020, Di et al. \cite{mobileApp} conducted a large-scale survey of dark patterns, targeting 240 popular mobile apps. Multiple researchers have analyzed the screenshots of pages within apps to investigate the presence of dark patterns. When dark patterns were identified, they were classified according to the five categories defined by Gray et al. \cite{darkSideUx}. They found that dark patterns were present in 95\% of the surveyed applications.

In 2020, Nouwens et al. \cite{dpGdpr} scraped 10,000 popular web pages in the U.K. and investigated 680 cookie consent designs to determine whether they contained dark patterns. In addition, they examined whether the collected cookie consent designs violated the regulations set by the European EU’s GDPR. They found that only 11.8\% met the minimal requirements regulated by EU law.

In 2023, Hidaka et al. \cite{linguistic} surveyed dark patterns by targeting 200 popular apps in Japan. They recorded the use of the apps in a video and investigated whether dark patterns existed by confirming the video. They found an average of 3.9 dark patterns per app.

\subsection{Dark Pattern Auto Detection}

Several studies have investigated the automatic detection of dark patterns.

In 2022, we adopted deep-learning-based text classification methods to present the baseline performance for the automatic detection of dark patterns on e-commerce sites \cite{dpInEc}. We used our new freely distributed text-based dataset to evaluate automatic dark pattern detection schemes, named the ”E-Commerce Dark Pattern Dataset,” which was created based on the original dark patterns collected by Mathur et al.  \cite{11kScale}. The RoBERTa model achieved an F1 score of 0.966.

In 2023, Mansur et al. \cite{aidui} proposed a framework called AidUI to detect and classify dark patterns. They constructed a Context DP dataset consisting of 162 webpages and 339 mobile screenshots to evaluate their framework. Using deep-learning-based models, they extracted user interface (UI) components and texts from screenshot images and then determined the existence of dark patterns. This model achieved an F1 score of 0.65.

\section{Explainable Dark Pattern Detection}\label{sec:method}
\subsection{Dark Pattern Auto Detection Model}

We constructed a detection model for dark patterns on e-commerce sites using fine-tuned transformer-based pre-trained language models,  including BERT \cite{bert} and RoBERTa \cite{roberta} which are the same as in our previous work \cite{dpInEc}.

Bidirectional Encoder Representations from Transformers (BERT) are language models with a structure that stacks multiple layers of the transformer encoder. BERT acquires linguistic knowledge through pretraining with a large amount of language data based on masked language model prediction (MLM) and next sentence prediction (NSP). Thus, transformer-based pretrained language models possess a high capability to represent contextual information in complex sentences.

\begin{figure}[t]
\includegraphics[width=\linewidth]{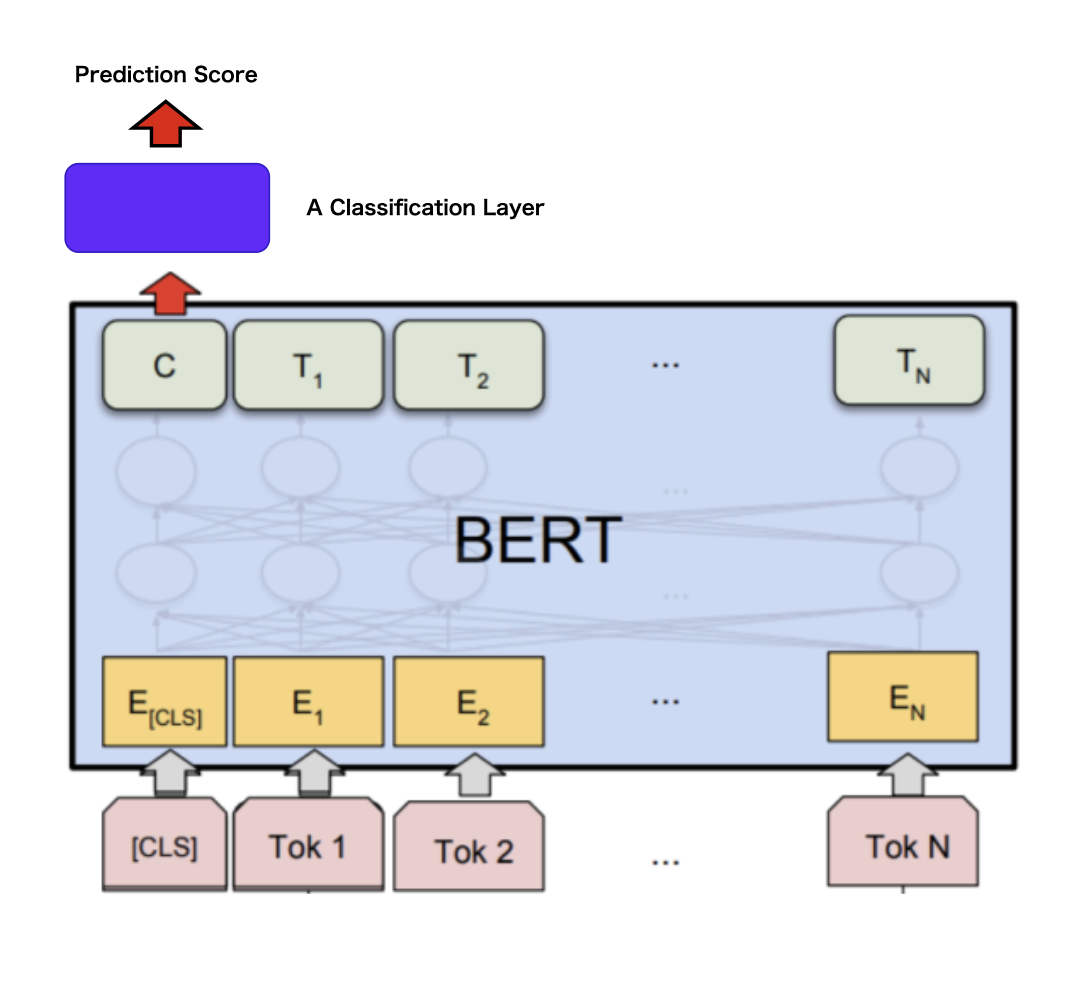}
\caption{Dark Pattern Auto Detection Model}
\label{bert}
\end{figure}

The automatic detection model for dark patterns using the transformer-based pretrained language models is shown in Fig. \ref{bert}. A classification layer was added to the BERT output layer to transform the CLS representation linearly. The output of the classification layer is trained to determine whether the input text is a Dark Pattern.

% https://app.diagrams.net/#G1eylaYLQbTa1enWNrJR6EiJF9r9XAzWqh

\subsection{Post-Hoc Interpretation of Dark Pattern Auto Detection Model}

In this section, we discuss methods for interpreting the automatic dark pattern detection model.

Local interpretable model-agnostic explanation (LIME) [11] is an algorithm that can interpret machine learning models without depending on their internal structure. LIME approximates the prediction results of the target model $f$ by training a distinct comprehensible surrogate model (e.g., Logistic Regression) model $g$. The contributions of the features in model $g$ explain $f$.

SHapley Additive exPlanations (SHAP) \cite{shap} is a coalition game theory method used to explain a model's prediction. SHAP samples permutations of the input features and aggregates the output prediction changes of model $f$. Then, SHAP calculates the contribution of each feature to the predictions.

\section{Experimental Settings}\label{sec:exp}
\subsection{Dataset}

We used the E-Commerce Dark Pattern Dataset \cite{dpInEc} to train a dark pattern auto detection model. The E-Commerce Dark Pattern Dataset consists of texts from e-commerce sites labeled as either dark or non-dark patterns, totaling 2,356 entries. Half of the dataset (1, 178 entries) consists of dark pattern texts, while the other half (1, 178 entries) consists of non-dark pattern texts. Dark pattern text is based on the text obtained from a large-scale survey by Mathur et al. \cite{11kScale}. On the other hand, Non-Dark Pattern Texts were acquired by scraping websites that did not have dark patterns from e-commerce sites accessed by Mathur et al. during their large-scale survey.

\subsection{Model Training and Evaluation}

We used the transformer library \footnote{https://huggingface.co/transformers/} to implement a dark pattern auto detection model using a transformer-based pretrained language model. For the construction, we applied BERT \cite{bert} and RoBERTa \cite{roberta} as backbone models. The AdamW \cite{adamw} optimizer and a linear learning rate scheduler were employed. To maximize the F1 score, we tuned the hyperparameters using a grid search. Table \ref{tab:hyper} shows the best hyperparameters for the adopted dark pattern auto detection models .

We evaluated the model performance using 5-fold cross-validation. The evaluation results are listed in Table \ref{tab:result}. The LIME and SHAP interpretation methods were applied to the model using $RoBERTa_{large}$, which exhibited the best performance.

\begin{table}[t]
	\centering
	\caption{Hyper Parametors}
	\def\arraystretch{1.5}
	\begin{tabular}{cccccc}  \hline
		Model  &  Batch Size & Learning Rate & Dropout rate & Epochs \\ \hline
		$\text{BERT}_{base}$ & 32 & 5e-5 & \multirow{4}{*}{0.1} & \multirow{4}{*}{5}  \\

            $\text{BERT}_{large}$ & 32 & 5e-5 &  &    \\
  
		$\text{RoBERTa}_{base}$ & 64 & 5e-5	& 	&   \\ 
  
		$\text{RoBERTa}_{large}$ & 32 & 3e-5 &  &    \\ 
	 \hline
	\end{tabular}
	\label{tab:hyper}
\end{table}

\begin{table}[t]
	\centering
	\caption{Performance Evaluation (Dark Pattern Auto Detection)}
	\def\arraystretch{1.5}
	\begin{tabular}{cccccc}  \hline
		Model  & Accuracy & AUC & F1-score  & Precision & Recall \\ \hline
		$\text{BERT}_{base}$ & 0.958 & 0.991 & 0.959 & 0.949 & $\mathbf{0.969}$  \\ 

            $\text{BERT}_{large}$ & 0.967 & 0.992 & 0.967 & 0.972 & 0.962  \\
  
		$\text{RoBERTa}_{base}$ & 0.965	& $\mathbf{0.992}$	& 0.965	& 0.970	& 0.960  \\ 
  
		$\text{RoBERTa}_{large}$ & $\mathbf{0.969}$ & 0.991 & $\mathbf{0.969}$ & $\mathbf{0.981}$ & 0.957  \\ 
	 \hline
	\end{tabular}
	\label{tab:result}
\end{table}

% TODO: Hyperparametors

\subsection{Local Interpretation by LIME}

We employed LIME to explain the results predicted by the dark pattern auto detection model. LIME was trained with all the dark pattern and non-dark pattern texts with the word vector. We set the parameters of LIME as a kernel width $\sigma=100$, number of neighborhood samples $S=25$, and number of features $K=10$.

After training, we calculated the model fidelity based on the accuracy of the prediction result by the LIME surrogate model $g$. The model fidelity was 0.96. 

\begin{figure}[b]
\includegraphics[width=\linewidth]{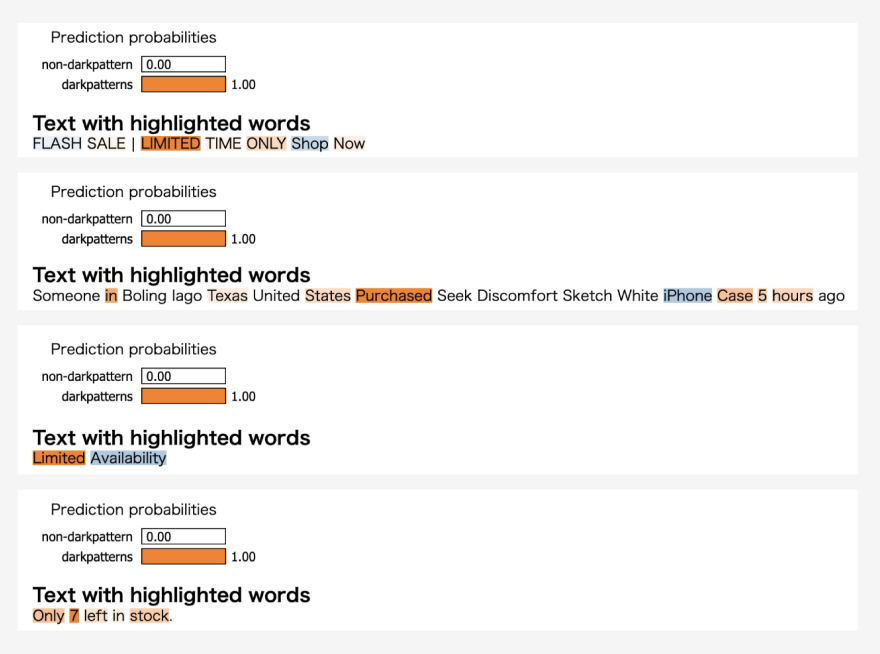}
\caption{Examples of LIME-based interpretation }
\label{lime}
\end{figure}

Fig. \ref{lime} shows examples of LIME-based interpretation for the texts corresponding to dark patterns, where the darker the orange color, the more likely it is to be a dark pattern. The results show that the model’s prediction of a dark pattern tends to be influenced by terms such as ”Limited,” ”Only,” and ”Purchased.” Dark patterns on e-commerce sites often pressure users by displaying limited stock or sale duration or by indicating popularity from other users to encourage purchases \cite{11kScale,darkpattern.org}. The tendencies observed in the extracted features aligned with the dark patterns commonly found on e-commerce sites. 

\subsection{Extraction of Features Influencing on Dark Patterns}

We applied SHAP to all dark pattern texts. We input the text and extracted the terms that influenced the dark pattern. SHAP outperforms LIME for global interpretations, i.e., it extracts terms that influence the prediction whether the  text is dark pattern entirely or not, of the model prediction. We calculated the average score ranged from 0 to 1, from SHAP and extracted words in descending order based on their average scores. Table \ref{tab:word} lists the words with high influence scores.

\begin{table}[t]
	\centering
	\caption{Terms extraction by SHAP}
	\def\arraystretch{1.5}
	\begin{tabular}{cccccc}  \hline
		  & Terms & Scores \\ \hline
1 & $\text{selling}$ & 0.769 \\
2 & $\text{yes}$ & 0.665 \\
3 & $\text{would}$ & 0.660 \\
4 & $\text{port}$ & 0.576 \\
5 & $\text{no}$ & 0.571 \\
6 & $\text{added}$ & 0.557 \\
7 & $\text{low}$ & 0.528 \\
8 & $\text{high}$ & 0.500 \\
9 & $\text{few}$ & 0.500 \\
10 & $\text{quantity}$ & 0.500 \\
11 & $\text{Compare}$ & 0.499 \\
12 & $\text{expire}$ & 0.496 \\
13 & $\text{last}$ & 0.490 \\
14 & $\text{limited}$ & 0.480 \\
15 & $\text{demand}$ & 0.477 \\
16 & $\text{risk}$ & 0.467 \\
17 & $\text{sell}$ & 0.451 \\
18 & $\text{purchased}$ & 0.437 \\
19 & $\text{already}$ & 0.427 \\
20 & $\text{only}$ & 0.419 \\
21 & $\text{bought}$ & 0.409 \\
22 & $\text{sold}$ & 0.386 \\
23 & $\text{less}$ & 0.384 \\
24 & $\text{withdraw}$ & 0.375 \\
25 & $\text{remaining}$ & 0.345 \\ \hline
	\end{tabular}
	\label{tab:word}
\end{table}

\section{Analysis of Terms Impacting on Dark Patterns}\label{sec:ana}

We extracted terms influencing dark patterns using a SHAP-based explanation. We applied SHAP to all dark patterned texts and calculated the SHAP scores. We then extracted terms with high SHAP scores and carefully categorized the extracted words in addition to reviewing the original sentences. We identify four categories of words that influence dark patterns on e-commerce sites.

\subsection{Fear of Missing Out}

The \textit{Fear of Missing Out} category refers to a term within a word affecting dark patterns, emphasizing to users that the product is about to sell out. E-commerce sites often display that there is limited stock left for a product, and sometimes this could potentially be intentional misinformation that prompts users to make a purchase. Words indicating limited stocks such as \textit{few}, \textit{only}, and \textit{last} influence the prediction of a dark pattern.

\subsection{Consensus on Popularity}

Several extracted words fall under the \textit{Consensus on Popularity} category. \textit{Consensus on Popularity} refers to terms like \textit{purchased}, \textit{demand}, and \textit{bought} that are used to indicate the purchasing status of other users. On e-commerce sites, by displaying statements like "32 people purchased the product" or "This product is in high demand," they influence users to purchase by indicating the product's popularity, which leverages the bandwagon effect \cite{bandwagon}.

\subsection{Sense of Urgent}

In the \textit{ Sense of Urgent } category, the terms indicating a ticking clock are highlighted. Various e-commerce platforms employ countdown UIs to showcase special deals or time-sensitive discounts, instilling a feeling of urgency in users and nudging them towards purchases. Words that imply haste, such as \textit{limited}, \textit{expire}, and \textit{hurry} influence the model's prediction of a dark pattern.

\subsection{Special Offer}

Some terms are categorized as \textit{ Special Offers }. These terms suggest \textit{Special Offers}, such as discounts or coupon distributions, exclusively for browsing users. Some e-commerce sites present deceptive special offers, displaying messages like "exclusive to specific users," but in reality, they show it to all users, leading them to purchase. Terms such as \textit{offers}, \textit{selling}, and \textit{exclusive} fall under the \textit{Special Offer} category.

\section{Conclusion}\label{sec:con}

In this study, we experimented with explaining why a given UI is a dark pattern by adopting explainable AI methods. Utilizing transformer-based pretrained language models, we trained a model for the automatic detection of dark patterns and applied LIME and SHAP. We analyzed the calculated explainability and extracted the terms that determined the presence of dark patterns. We identified the extracted terms under four categories: fear of missing out, consensus on popularity, sense of urgency, and special officers.

\section*{Acknowledgments}

Part of this work was funded by the LINE Corporation.

\bibliographystyle{IEEEtran} %引用された順番に出力
\bibliography{why_darkpattern} %bibファイルの.bibの前の部分

\end{document}